\documentstyle[12pt]{article}

\newcommand{\be}{\begin{equation}}
\newcommand{\ee}{\end{equation}}
\newcommand{\ba}{\begin{eqnarray}}
\newcommand{\ea}{\end{eqnarray}}
\newcommand{\bn}{\begin{eqnarray}}
\newcommand{\en}{\end{eqnarray}}
\newcommand{\p}{\partial}
\def\l{\label}
\def\ni{\noindent}

\newcommand{\ffi} {\phi^i}
\newcommand{\fj} {\phi^j}
\newcommand{\fk} {\phi^k}
\newcommand{\ix} {\int\!\!d^2x\;}
\newcommand{\intt} {\int\!\! }

\newcommand{\dpl} {\partial_+}
\newcommand{\dmi} {\partial_-}

\def\p {C}

\begin{document}

\begin{center}

\noindent{\large\bf
{\Huge\bf New approaches in ${\cal W}$-gravities}
}\vspace{6mm}

\noindent
{\Large E. M. C. Abreu$^{a}$, A. de Souza Dutra$^{a}$ and C. Wotzasek$^{a,b}$
}\vspace{3mm}

\noindent

\begin{large}
$^a$
Departamento de F\'{\i}sica e Qu\'{\i}mica, UNESP/Guaratinguet\'a,  \\
Av. Ariberto Pereira da Cunha 333, Guaratinguet\'a, \\ 12516-410, S\~ao Paulo,
SP, Brazil. \\
{\sf E-mail: everton@feg.unesp.br and dutra@feg.unesp.br}\\
$^b$
Instituto de F\'{\i}sica, Universidade Federal do Rio de Janeiro,\\
Caixa Postal 68528, 
21945-970, Rio de Janeiro, RJ, Brazil. \\
{\sf E-mail: clovis@if.ufrj.br and wotzasek@feg.unesp.br} \\
\end{large}
\vspace{4mm}

\today

\vspace{7mm}

\end{center}

\begin{abstract}

\ni  We have devoted an effort to study some nonlinear actions, characteristics
of the ${\cal W}$-theories, in the framework of the soldering formalism. We have disclosed interesting new results concerning the embedding of
the original chiral ${\cal W}$-particles in different metrical spaces in the final soldered action, i.e.,
the metric gets modified by the soldering interference process.  The results are presented in a weak field approximation for the ${\cal W}_N$ case when $N\geq 3$ and also in an exact way for ${\cal W}_2$.
We have promoted a generalization of the interference phenomena to
${\cal W}_N$-theories of different chiralities and shown that the geometrical features
introduced can yield a new understanding about the interference formalism in quantum field theories.
\end{abstract}

PACS: 03.70.+k, 11.10.Ef,11.15.-q

\newpage

\section{Introduction}

This paper is devoted to study the effects of interference between the chiral modes carrying
a representation of conformal spins of order higher than two in the context of the soldering
formalism.  These modes are described by chiral ${\cal W}_N$ gravities for
$N\geq 2$. This study is a natural and deeper extension of
\cite{adw} where the soldering of two chiral ${\cal W}_2$ particles (analogous to the Siegel
particles \cite{siegel}) was shown to produce the action for
a non-chiral 2D scalar field coupled to a gravitational background.  It is therefore natural
to ask for the possibility of soldering chiral modes carrying
representations of higher conformal spins.

There are two main reasons for studying extended symmetries in conformal
field theory \cite{bs}. Certain applications of conformal field
theory either in string theory or statistical mechanics, require some additional
symmetry besides conformal invariance. Moreover
extended symmetries can help in the analysis of a large
class of conformal field theories (called rational conformal field theories)
and to classify certain types of conformal field theories.

Conformal invariance in two dimensions is a powerful
symmetry that allows certain 2D-dimensional quantum field theories to be solved exactly.
Conformal field theories have found remarkable applications in
string theory (see \cite{gsw}). This study together with the investigation of critical phenomena in
statistical mechanics (see \cite{isz}, for selected reprints),
has produced a large-scale study of conformal field theories in
recent years.
For example, the study of so-called perturbed
conformal field theories
has given rise to surprising new results for certain massive integrable
quantum field theories \cite{zab}.

Additional motivation
for a detailed study of the infinite dinebsional algebraic structure of conformal field theories comes from
the study of two-dimensional gravity, that relies heavily on conformal field
theory techniques, and from two-dimensional topological
quantum field theories.
Infinite-dimensional symmetry algebras are known to play a central r\^ole in 2D physics.
There is an intrinsic connection of these algebras with two-dimensional
gauge theories or string theories, the most important example being the symmetry algebra of two dimensional
conformal field theories, i.e., the Virasoro algebra.

The Virasoro algebra admits higher-spin extension, known as ${\cal W}_N$-algebras, containing
generators with conformal spins $2,3,4,\ldots,N$ \cite{zam,fl}.
A ${\cal W}$-algebra is an extended conformal algebra that satisfies
the Jacobi identities and contains the Virassoro algebra as a subalgebra \cite{bs,hull1}.
${\cal W}$-algebras are infinite-dimensional symmetry algebras with the restriction that at least one
of the generating currents has spin $s\geq 2$. 
This algebra is generated by a set of chiral currents which are our main interest in the present stydy.

In addition to the above motivation for the present study in extended conformal symmetry we would like to mention
that in perturbed conformal
field theories, the presence of ${\cal W}$-symmetries in the original
conformal field theory may lead to additional integrals of motion in
the perturbed theory \cite{zaaa} and for the relation with topological
field theories, $N=2$ extended superconformal symmetry is essential
\cite{wiaa,wiab,ey}.
Extended symmetries also appear to be particularly important for the coupling
of conformal field theory `matter systems' to two-dimensional gravity \cite{kpz}.
Classical and quantum ${\cal W}$-gravity, in particular ${\cal W}_3$ gravity, have
recently been studied by various groups \cite{bs}.

In order to apply the soldering formalism to the opposite chiral-${\cal W}$ aspects we need an explicite field theoretical realization of these algebras. Realizations of the chiral
${\cal W}_N$-algebras have been constructed, for example in terms of $(N-1)$ free bosons via the Miura
transformations \cite{fl}.
Other realizations, using a generalizing Sugawara construction,
have also been given \cite {bbss}.  This seems important since tentative extensions of string-theory based on extra bosonic symmetry (${\cal W}$-symmetry) on the worldsheet have been proposed and are called ${\cal W}$-strings \cite{bge,ddr,prstb} which are higher spin generalizations of ordinary string theories, such that the string coordinates are not only coupled to the world-sheet metric but also to a set of higher spin world-sheet gauge fields (for a review see \cite{hull5}).  Since ordinary string theory can be considered as a gauge theory based on a the Virassoro algebra, one can analogously define a ${\cal W}$-string theory as a gauge theory based on a ${\cal W}$-algebra \cite{zam,fz,fl} (or any other higher spin conformally extended algebra \cite{hull5}).
The bosonic representations of the chiral algebras will be the starting
point for our application of the soldering formalism.

Recently, there has been a great improvement to solder together distinct manifestations of chiral and
duality symmetries [23-32].
The procedure leads to new physical results that includes the idea of
interference effect.  
The soldering formalism was introduced in \cite{ms,wzw} to solder together two chiral scalars by introducing a non-dynamical gauge field to remove the degree of freedom that obstructs the vector gauge invariance.  This is connected via chiral bosonization to the necessity that one has to have more than the direct sum of two fermions representations of the Kac-Moody algebra to describe a Dirac fermion.  In other words we can say that the equality for the weights in the two representations is physically connected with the necessity to abandon one of the two separate chiral symmetries, and accept that only vector gauge symmetry should be maintained.  Besides, being just an auxiliary field, it may posteriorly be eliminated (integrated) in favor of the physically relevant quantities.  This restriction will force the two independent chiral representations to belong to the same multiplet, effectively soldering them together.  This is the main motivation for the introduction of the soldering field which permeates to the case of higher conformal spin currents.

In section 2 we make a review of the ${\cal W}$-theories.  The soldering formalism is shortly depicted in section 3.  The fusion of chiral ${\cal W}_N$-particles is accomplished in section 4.  Conclusions and final remarks are shown in section 5.

\section{The ${\cal W}$ Gravities}

In order to make this work self-contained, in this section we will make a brief review of the ${\cal W}$ realizations and gauging following closely the references \cite{hull1,hull4,hull,hull2}.

\subsection{The ${\cal W}$ Algebra}

Let us consider Lie algebras under Poisson brackets,
with the generators $t_a$ labeled by an index $a$ (which may be of infinite range)
\be
\label{lie}
\lbrace t_a,t_b\rbrace ={f_{ab}}^ct_c +c_{ab}
\ee
where ${f_{ab}}^c$ are the structure constants and $c_{ab}$ are constants
defining the central extension of the algebra. However, for many ${\cal W}$-algebras, the
Poisson bracket structure give a result non-linear in the generators
\be
\label{nonlie}
\lbrace t_a,t_b\rbrace ={f_{ab}}^ct_c +c_{ab}+{g_{ab}}^{cd}t_ct_d+\dots =F_{ab}(t_c)
\ee
and the algebra is said to close in the sense that the right-hand-side is a
function of the generators.
Most of the ${\cal W}$-algebras that are generated by a finite number of currents,
are non-linear algebras of
this type.
At first sight, it appears that there might be a problem in
trying to realize a non-linear algebra in a field theory, as symmetry
algebras are usually Lie algebras. However, as will be seen, a non-linear
algebra can be realized as a symmetry algebra for which the structure constants
are replaced by field-dependent quantities.

A field theory with action $S_0$ and conserved symmetric tensor currents
$T_{\mu \nu},W^A_{\mu_1 \mu_2 \dots \mu _{s_A}}$
(where $A=1,2,\dots $ labels the currents, which have spin $s_A$) will be
invariant under global symmetries  with constant parameters $k^\mu, \lambda_A^{\mu_1 \mu_2 \dots \mu _{s_A-1}}$ (translations and ${\cal W}$-translations) generated by the Noether charges $P_\mu, Q^A_{\mu_1 \mu_2 \dots \mu _{s_A-1}}$
(momentum and ${\cal W}$-momentum) given by  $P_\mu=\int t dx^0T_{0\mu}$ and  $ Q^A_{\mu_1 \mu_2 \dots \mu _{s_A-1}}=\intt dx^0
W^A_{\mu_1 \mu_2 \dots \mu _{s_A-1}0}$.
If   the currents are traceless, then the theory
is invariant under an infinite dimensional extended conformal
symmetry. The parameters $ \lambda_A^{\mu_1 \mu_2 \dots \mu _{s_A-1}}$ are then traceless and the corresponding transformations will be symmetries if the parameters are not constant but satisfy the conditions that the trace-free parts of $\partial ^{(\nu}k^{\mu)}, \partial ^{(\nu}
 \lambda_A^{\mu_1 \mu_2 \dots \mu _{s_A-1})}$ are zero.  This implies that
$\partial_{\mp}k^ \pm=0$ and $ \partial_{\mp}
\lambda _A^{\pm \pm \dots \pm} = 0$ so that the parameters are semi-local, $k^\pm = k^\pm(x^\pm)$ and $\lambda_A^{\pm \pm \dots \pm} =  \lambda _A^{\pm \pm \dots \pm}(x^\pm)$
which are the parameters of conformal and ${\cal W}$-conformal transformations.

The soldering formalism, to be developed in the next section, needs to promote the lift of these global symmetries
to their gauge invariant version.
The global symmetries corresponding to the currents $T_{\mu \nu},W^A_{\mu_1
\mu_2 \dots \mu _{s_A}}$ are promoted to local ones by coupling to
the ${\cal W}$-gravity gauge fields
$h^{\mu \nu}, B_A^{\mu_1
\mu_2 \dots \mu _{s_A}}$ which are symmetric tensors transforming, to lowest order in the gauge fields, as
\be
\label{traaaa}
\delta h^{\mu \nu}=\partial ^{(\nu}k^{\mu)}+\dots ,\qquad
\delta B_A^{\mu_1
\mu_2 \dots \mu _s}=\partial ^{(\nu}
\lambda_A ^{\mu_1 \mu_2 \dots \mu _{s-1})}+\dots,
\ee
The action is given by the Noether
coupling
\be
\label{snoth}
S=S_0+ \ix \left( h^{\mu \nu}T_{\mu \nu}+ B_A^{\mu_1
\mu_2 \dots \mu _{s_A}}W^A_{\mu_1
\mu_2 \dots \mu _{s_A}}\right) + \dots
\ee
plus terms non-linear in the gauge fields. If the currents
$T_{\mu \nu},W^A_{\mu_1
\mu_2 \dots \mu _s}$
are traceless, in the sense that there is an extended conformal symmetry, then the traces of
the gauge fields decouple and the theory is invariant under Weyl and ${\cal W}$-Weyl transformations given to lowest order in the gauge fields by
\be
\label{wweyl}
\delta h^{\mu \nu}= \Omega \eta  ^{\mu \nu}+\dots ,
\qquad
\delta B_A^{\mu_1
\mu_2 \dots \mu _s}=\Omega_A ^{(\mu_1
\mu_2 \dots \mu _{s-2}}\eta  ^{\mu _{s-1}\mu_s)}+\dots
\ee
where $\Omega(x^\nu)$, $\Omega_A^{ \mu_1
\mu_2 \dots \mu _{s-2}}(x^\nu)$ are the local parameters.
This defines the linearized coupling to ${\cal W}$-gravity. The full non-linear
theory is in general non-polynomial in the gauge fields of spins 2 and higher, which makes matter harder.
The non-linear theory can be constructed to any given order using the Noether
method, but to obtain the full non-linear structure requires a deeper
understanding of the geometry underlying ${\cal W}$-gravity which is beyond the scope of this study.

\subsection{The ${\cal W}$ Field Theory}

Consider a field theory in flat Minkowski space with metric $\eta_{\mu \nu}$
and coordinates $x^0,x^1$. The stress-energy tensor is a symmetric tensor
$T_{\mu \nu}$ which, for a translation-invariant theory, satisfies the conservation law
\be
\label{tcon}
\partial ^\mu
T_{\mu \nu}=0
\ee
A spin-$s$ current in flat two-dimensional space is a rank-$s$ symmetric
tensor $W_{\mu_1 \mu_2 \dots \mu _s}$ and will be conserved if
\be
\label{wcon} \partial ^{\mu_1} W_{\mu_1 \mu_2 \dots \mu _s}=0
\ee

Recall that, in two dimensions, a rank-2 tensor can be decomposed as, e.g., $V_{\mu \nu}=V_{(\mu \nu)}+V\epsilon _{\mu \nu}$ where $V=\frac 12 \epsilon ^{\mu
\nu}V_{\mu \nu}$. Thus without loss of generality, all the conserved
currents of a given theory can be taken to be symmetric tensors. A rank-$s$
symmetric tensor transforms as the spin-$s$ representation of the two-dimensional Lorentz group.

A theory is conformally invariant if the stress tensor is traceless, ${T_\mu}^\mu=0$. Introducing null coordinates $x^\pm={1 \over \sqrt 2} (x^0 \pm x^1)$, the tracelessness
condition becomes $T_{+-}=0$ and (\ref{tcon}) then implies that the remaining
components $T_{\pm\pm}$ satisfy
\be
\label{rert}\dpl T_{--}=0, \qquad \dmi T_{++}=0
\ee

If a spin-$s$ current $W_{\mu_1 \mu_2 \dots \mu _s}$ is traceless, it will
have only two non-vanishing components, $W_{++ \dots  +}$ and $W_{--\dots -}$. The conservation condition (\ref{wcon}) then implies that
\be
\label{dfg} \dmi W_{++ \dots  +}=0, \qquad \dpl W_{--\dots -}=0
\ee
so that $W_{++ \dots  +}=W_{++ \dots  +}(x^+)$
and $W_{--\dots -}=W_{--\dots -}(x^-)$ are right- and left-moving chiral
currents, respectively.  For a given conformal field theory, the set of all right-moving chiral
currents generate a closed current algebra, the right-moving chiral algebra,
and similarly for left-movers. The right and left chiral algebras are
examples of ${\cal W}$-algebras that are main components of our construction here.

Consider a set {\cal S} of right-moving chiral currents $T(x^+)=T_{++}(x^+),W(x^+),\dots$ of spins $2,s_W,\dots$.  The main example that will be of interest here is that in which
the currents arise from some field theory and the bracket is the
Poisson bracket in a canonical formalism in which $x^-$ is regarded as the
time variable. The current $T$ satisfies the conformal algebra if
\be
\label{con}
\lbrace T(x^+),T(y^+)\rbrace = -\,
\delta ^\prime (x^+-y^+)[T(x^+)+T(y^+)]
\ee
in which case its modes $L_n$ generate the Virasoro algebra.
A current $W$ is said to be primary of spin $s_W$ if
\be
\label{prim}
\lbrace T(x^+),W(y^+)\rbrace = -\,
\delta ^\prime (x^+-y^+)[W(x^+)+(s_W-1)W(y^+)]
\ee
The set ${\cal S}$ of currents will generate a ${\cal W}$-algebra if the bracket of any
two currents gives a function of currents in ${\cal S}$ and if the bracket satisfies the Jacobi identities.

Consider first the  case in which there are just two currents, $T$ and $W$, where $W$ is  primary of spin $s=s_W$ and that the $\lbrace W,W\rbrace $ bracket takes the form
\be
\label{walg}
\lbrace W(x^+),W(y^+)\rbrace = -2
\kappa \delta ^\prime
(x^+-y^+)[\Lambda(x^+)+\Lambda(y^+)]
\ee
for some $\Lambda$, where $\kappa$ is a constant. If the algebra is to
close, the current $\Lambda$ must be a function of the currents $T,W$ and
their derivatives. If $s=3$, then $\Lambda$ is a spin-four current and the
Jacobi identities are satisfied if
\be
\label{lamis}
\Lambda=TT
\ee
The algebra then closes non-linearly.  Notice that in the limit $\kappa \rightarrow 0$,
this contracts to a linear algebra.
For $s>3$,  the algebra will again close and satisfy the Jacobi identities
if $\Lambda $  depends on $T,W$ but not on their derivatives.
If $s$ is even, the most general such $\Lambda $ is  of the form
\be
\label{lop}
\Lambda = \alpha T^{s-1} +\beta WT^{s/2-1}
\ee
for some constants $\alpha ,\beta$,
while if $s$ is  odd, such a $\Lambda $ must
be of the form (\ref{lop}) with $\beta=0$.
The   algebra given by (\ref{con}), (\ref{prim}), (\ref{walg}) and (\ref{lop}) is the algebra $W_{s/s-2}$ of ref. \cite{hull2}.

A large number of ${\cal W}$-algebras are now known. The ${\cal W}_N$ algebra \cite{bg} has currents of spins $2,3,\ldots ,N$ (so that ${\cal W}_2$ is the Virasoro algebra), the
${\cal W}_\infty $ \cite{ib,prs}
algebra has currents of spins $2,3,\dots ,\infty$ while the
${\cal W}_{1+\infty} $ algebra \cite{prs} has
currents of spins $1,2,3,\dots ,\infty$.

Consider a theory of $D$ free scalar fields $\ffi$ $(i=1,\dots,D)$ with action
\be
\label{free}S_{0}=\ix \partial_{+}\phi^{i}\partial_{-}\phi^{i}
\ee
where the two-dimensional space-time has null coordinates $x^\mu=(x^+,x^-)$ which are related to the usual Cartesian coordinates by $x^\pm ={1 \over \sqrt 2}( x^0 \pm x^1)$.  The stress-energy tensor
\be
\label{tis}
T_{++}={1 \over 2}
\partial_{+}\phi^{i}\partial_{+}\phi^{i}
\ee
is conserved, $\partial_{-}T_{++}=0$, and  generates the   Poisson bracket algebra (\ref{con}) (in a canonical treatment regarding $x^-$ as time \cite{wzw}) which is the conformal algebra with vanishing central charge.  For any   rank-$s$ constant symmetric tensor $d_{i_1 i_2 \dots i_s}$ one can construct a current
\be
\label{freecur}
W_{++\dots +}= {1 \over s}
d_{i_1 i_2 \dots i_s}\partial_{+}\phi^{i_1}\partial_{+}\phi^{i_2}\dots\partial_{+}\phi^{i_s}
\ee
which is conserved, $\partial _-W=0$, and which is a spin-$s$ classical
primary field - its Poisson bracket with $T$ is given by (\ref{prim}).  The Poisson bracket
of two $W$'s is (\ref{walg}), where $\Lambda $ is given by
\be
\label{ewrte}\Lambda ={1 \over 4  \kappa}
{d_{i \dots j}}^{  m}
d_{k \dots lm} \dpl \ffi  \dots \dpl \fj \dpl \fk \dots \dpl \phi ^l
\ee
(the indices $i,j,\dots$ are raised and lowered with the flat metric $\delta_{ij}$).

Consider first the case $s=3$. In general, closing the algebra generated
by $T,W$ will lead to an infinite sequence of currents $(T,W, \Lambda ,\dots)$.
However, if  $\Lambda = T^2 $, for some constant $\kappa$, then the algebra closes non-linearly   on $T$ and $W$, to give the so-called classical ${\cal W}_3$-algebra depicted above.

In \cite{hull}, it was shown that for any number $D$ of bosons, the
necessary and sufficient condition for (\ref{lamis}) to be satisfied and hence
for the classical ${\cal W}_3$ algebra to be generated is that the ``structure constants" $d_{ijk}$ satisfy
\be
\label{did}
d^{\ \ \ m}_{(ij}d_{k)lm}=\kappa \delta_{(ij}\delta_{k)l}\;\;.
\ee

\noindent This rather striking algebraic constraint has an interesting algebraic
interpretation\footnote{This identity has in fact occurred at least once
before in the physics literature, in the study of five-dimensional
supergravity theories \cite{gst}.}. It implies that the $d_{ijk}$
are   the structure constants for a Jordan algebra (of degree $3$)
\cite{ljr} which is a commutative algebra for which  (\ref{did}) plays the r\^ole of
the Jacobi identities. Moreover, the set of all such algebras has been
classified \cite{rds}, allowing one to write down the general solution to (\ref{did})
\cite{ljr}. In particular, (\ref{did}) has a solution for any number $D$ of
bosons. Examples of solutions to (\ref{did}) are given for $D=1$ by
$d_{111}=\kappa$ and  for $D=8$ by taking  $d_{ijk}$ proportional to the
$d$-symbol for the group $SU(3)$ \cite{wzw}. For $D=2$, the construction of
\cite{fz} gives a solution of (\ref{did}) in which the only non-vanishing components
of $d_{ijk} $ are given by $d_{112}=-\kappa $ and $d_{222}=\kappa$, together
with those related to these by symmetry. The conserved currents $T,W$ correspond to the invariance of the free action $S_0$ under the conformal symmetries
\be
\label{inv}
\delta\phi^{i}=k_{-}\partial_{+}\phi^{i}+\lambda_{--}d^{i}_{\
jk}\partial _{+}\phi^{j}\partial_{+}\phi^{k}
\ee

\ni where the parameters satisfy
\be
\partial_{-}k_{-}=0,\qquad \partial_{-}\lambda_{--}=0.
\ee

Symmetries of this kind whose parameters are only functions of $x^+$ (or
only of $x^-$)  will be referred to here as semi-local.
The   symmetry   algebra closes to give
\be
\left\lbrack\delta_{k_{1}}+\delta_{\lambda_{1}},\delta_{k_{2}}+\delta_{\lambda_{2}}\right\rbrack=\delta_{k_{3}}+\delta_{\lambda_{3}}
\ee
where
\bn
k_{3}&=&\left\lbrack k_{2}\partial_{+}k_{1}+4 \kappa (\lambda
_{2}\partial_{+}\lambda_{1})T_{++}\right\rbrack\ -\ (1\leftrightarrow2) \nonumber \\
\lambda_{3}&=&\left\lbrack2\lambda_{2}\partial_{+}k_{1}+k_{2}\partial_{+}\lambda_{1}\right\rbrack\ -\ (1\leftrightarrow2)\;\;.
\en

In particular, the commutator of two $\lambda$ transformations is a
field-dependent $k$-transfor-

\ni mation, which is precisely the transformation
generated by the spin four current $\Lambda=TT$.
The gauge algebra structure \lq constants'
are not constant but depend on the fields $\phi$ through the current $T$,
reflecting the $TT$ term in the current algebra.

To gauge these symmetries to a local ${\cal W}$-diffeomorphism, the spin two and three conformal Noether currents above are introduced in the model with the correspondent Lagrange multiplier fields $h_{--}$ (the graviton) and $B_{---}$ (the ${\cal W}$-graviton) leading to the Lagrangian of the right ${\cal W}_3$
model as,
\be
\l{oito}
{\cal L} \,=\, \partial_+ \phi^i \partial_- \phi^i\,+\,{\frac{h_{--} }{2}}
\partial_+ \phi^i \partial_+ \phi^i \,+\,\,{\frac{B_{---} }{3}} d_{ijk}
\partial_+ \phi^i \partial_+ \phi^j \partial_+ \phi^k ,
\ee

\ni which is the result of the Noether couplings.  Notice that the free action
is already invariant under ``right-moving" transformation.

It is well known that this model is invariant under the transformations (\ref
{inv}) and together with the following symmetries
\begin{eqnarray}
\delta h_{--} &=& \partial_{-}k_{-}+k_{-}\partial_{+}h_{--}-h_{--}\partial_{+}k_{-}\,+\, \, 2 \kappa \left(\lambda_{--}\partial_{+}B_{---}-B_{---}\partial_{+}\lambda
_{--}\right)T_{++}  \nonumber \\
\delta B_{---} &=& \partial_{-}\lambda_{--}+2\lambda_{--}\partial_{+}
h_{--}-h_{--}\partial _{+}\lambda_{--}\,-\,2B_{---}\partial_{+}k_{-}+k_{-}\partial_{+}B_{---}\;\;, \nonumber \\
& &
\end{eqnarray}
has extended the original theory to a ${\cal W}$-gravity, so that the
original semi-local conformal symmetries were promoted to a local ${\cal W}$%
-diffeomorphism. In a similar way we may also gauge the left-handed ${\cal W}
$-algebra generated by $T_{--}$ and $W_{---}$ with analogous definitions and
results. The terms with
$H_{--}$ and $k_-$ are analog to the gauging of the right-handed Virasoro algebra.
Hence we can see expressions similar to the two-dimensional gravity in the
chiral gauge \cite{polyakov} or from Siegel's analysis for the chiral boson \cite{siegel}.

The situation is similar for $s>3$. The algebra will close, i.e., (\ref{lop}) will be
satisfied, if the $d$-tensor in (\ref{freecur}) satisfies a quadratic constraint
\cite{hull2} and again this constraint has an algebraic interpretation \cite{hull2}.
The $k$ and $\lambda$-transformations become
\be
\delta\phi^{i}=k\partial_{+}\phi^{i}+\lambda d^{i}_{\
i_{1}...i_{s-1}}\partial _{+}\phi^{i_{1}}...\partial_{+}\phi^{i_{s-1}}
\ee
\ni where the parameters  satisfy $\partial_{-}k=0$, $ \partial_{-}\lambda=0$. The symmetry algebra again has field dependent structure \lq constants'.  More generally, any set of constant symmetric tensors $d^A_{ij \dots k}$ labeled by some index $A$ can be used to construct a set of conserved currents
\be
W^A_{++\dots +}=d^A_{ij \dots k}\partial_{+}\phi^{i}\partial_{+}\phi^j\dots
\partial_{+}\phi^k
\ee
\ni which are classical primary fields, i.e., their Poisson bracket with $T$ is (\ref{prim}).
The current algebra will close if the $d^A$ tensors satisfy certain algebraic
constraints and the Jacobi identities will automatically be satisfied as the
algebra occurs as a symmetry algebra. In this way, a large class of classical
${\cal W}$-algebras can be constructed by seeking $d^A$-tensors satisfying the
appropriate constraints.  $D$ boson realizations of the ${\cal W}_N$ algebras were constructed in this way in ref. \cite{hull2}, where it was shown that the ${\cal W}_N$ $d$-tensor constraints had an interpretation in terms of Jordan algebras of degree $N$, and this again
allowed the explicit construction of solutions to the $d$-tensor constraints.
These realizations of classical $c=0$ algebras can be generalized to ones
with $c>0$ by introducing a background charge $a_i$, so that the stress tensor
becomes $T=  \dpl \ffi \dpl \ffi + a_i \dpl ^2 \ffi$ and adding  appropriate higher derivative terms (i.e., ones involving $\dpl^m \ffi$ for $m>1$) to the other currents.
The classical central charge becomes $c=a^2/12$, and for the $N-1$ boson realization
of ${\cal W}_N$, the structure of the higher derivative terms in the currents
$W_n$ can be derived using Miura transform methods \cite{bg,fl}.

Another important realization of classical ${\cal W}$-algebras is as the Casimir
algebra  of   Wess-Zumino-Witten (WZW) models \cite{hull4}. For the WZW model
corresponding to a group $G$, the Lie-algebra valued currents $J_+=g^{-1}\dpl g$ generate a
Kac-Moody algebra and are (classical) primary with respect to the Sugawara
stress-tensor $T= {1 \over 2} tr (J_+J_+)$.  Similarly, the higher order Casimirs allow a generalized Sugawara construction of higher spin currents $tr(J_+^n)$.
For example, for $G=SU(N)$  the set of currents $W_n= {1 \over n } tr(J_+^n)$ for $n=2,3,\dots ,N$ generate a closed algebra which is a classical $W_N$ algebra \cite{hull4}; similar results hold  for other groups.

Quantum mechanically, however, the Sugawara expressions for the currents  need
normal ordering and must be rescaled \cite{bbss,jt}. For example, in the case of $SU(3)$, the quantum Casimir algebra  leads to a closed ${\cal W}$-algebra (after a certain
truncation) only in the case in which the Kac-Moody algebra is of level one \cite{bbss}.

${\cal W}$-algebras also arise as symmetry algebras of many other field theories,
including Toda-theories \cite{bg}, free-fermion theories \cite{hull4} and non-linear
sigma-models \cite{hull,hp}, giving corresponding realizations of ${\cal W}$-algebras.

\section{The interference of the chiral ${\cal W}$-theories}

\subsection{${\cal W}_2$-gravity in the weak field approximation}

Let us analyze the ${\cal W}_2$ model for the right-handed chirality,
which is obtained from (\ref{oito}) by making $d_{ijk}\to 0$, i.e.,
\bn
\l{oito2}
{\cal L}_+^0 &=& \partial_+ \phi^i \partial_- \phi^i\,+\,{\frac{h_{--} }{2}}
\partial_+ \phi^i \partial_+ \phi^i \;\;,
\en
The soldering transformation to be gauged, as depicted in the last section, is
\begin{equation}
\phi^i \rightarrow \phi^i \,+\,\alpha^i\;\;,
\end{equation}
where $\alpha^i$ is the semi-local gauge parameter.

The corresponding variation of the model under this transformation is,
\begin{equation}
\delta {\cal L}_+^0\,=\,J^+_i \partial_+ \alpha^i\;\;,
\end{equation}
where $J^+_i$ is the left Noether current given by
\begin{equation}
\l{37}
J^+_i\,=\,2\,\partial_- \phi^i\,+\,h_{--} \partial_+ \phi^i
\end{equation}

Following the soldering algorithm and computing only the final steps, we have after two iterations that
\begin{equation}
\delta {\cal L}^2_+\,=\,-\,2\,A_+^i \delta A_-^i\;\;,
\end{equation}
where $A^i_{\pm}$ are the soldering fields.

For the left chirality we can write
\be
{\cal L}^0_- = \partial_+ \rho^i \partial_- \rho^i\,+\,{\frac{h_{++} }{2}}
\partial_- \rho^i \partial_- \rho^i\;\;.
\ee

\ni Analogously, the variation of the model under this is
\begin{equation}
\rho^i \rightarrow \rho^i \,+\,\alpha^i
\end{equation}
and
\begin{equation}
\delta {\cal L}_-^0\,=\,J^-_i \partial_- \alpha^i\;\;,
\end{equation}
where $J^-_i$ is the right Noether current
\begin{equation}
\l{42}
J^-_i\,=\,2\,\partial_+ \rho^i\,+\,h_{++} \partial_- \rho^i\;\;.
\end{equation}
Again, after two iterations we have that
\begin{equation}
\delta {\cal L}^2_-\,=\,-\,2\,A_-^i \delta A_+^i\;\;.
\end{equation}
We can see easily that the final soldered action is
\begin{equation}
{\cal L}_{FINAL}\,=\,{\cal L}^2_+\,+\,{\cal L}^2_-\,+\,2\,A^i_-\,A^i_+
\end{equation}
which has the desired vectorial gauge invariance, i.e., $\delta {\cal L}\,=\,0$, as can be easily checked.  Substituting all the ${\cal L}_\pm^{(N)}$ computed before we can write
explicitly the final form of the action as being
\bn
\label{lag}
{\cal L}_{FINAL}\,&=&\,{\cal L}_+^0\,+\,{\cal L}_-^0\,-\,A^i_+\,J^+_
i\,-\,A^i_-\,J^-_ i\,  +\,{\frac{h_{--} }{2}}\,(A^i_+)^2\,+\,{\frac{h_{++} }{2}%
}\,(A^i_-)^2\,+\,2\,A^i_-\,A^i_+ \nonumber \\
& &
\en

Next, by solving the equations of motion for the soldering fields we have, 
\be
A^i_+\,=\,{\frac{1 }{2}} J^-_i\,-\,{\frac{1 }{2}}h_{++}\,A^i_-\,
\l{eqmova}
\ee
\be
A^i_-\,=\,{\frac{1 }{2}} J^+_i\,-\,{\frac{1 }{2}}h_{--}\,A^i_+\,
\l{eqmovb}
\ee
and these fields can be eliminated in favor of the other variables.

Susbstituting the $A^i_\pm$ defined in (\ref{eqmovb}) in (\ref{eqmovb}) and solving the sistem iteratively we obtain,
\bn
A_+&=&{1 \over 2}\,J^-\,-\,{1\over4}h_{++}\,J^+\,+\,h^2\,A_+ \nonumber \\
&=&A^{(0)}_+\,+\,h^2\,A_+\;\;,
\en
now substituting the second equation in the first and so on we have
\bn
A_+&=&A^{(0)}_+\,+\,h^2\,[A^{(0)}_+\,+\,A_+] \nonumber \\
&=&A^{(0)}_+\,+\,h^2A^{(0)}_+\,+\,h^4\,[A^{(0)}_+\,+\,A_+] \nonumber \\
&=&A^{(0)}_+\,+\,h^2A^{(0)}_+\,+\,h^4\,A^{(0)}\,+\,h^6\,A^{(0)}_+\,+\,\ldots \nonumber \\
&=&[1\,+\,h^2\,+\,h^4\,+\,h^6\,+\,\ldots]\,A^{(0)}_+ \nonumber \\
&=&f_\infty\,A^{(0)}
\en
where
\be
f_\infty\,=\,1\,+\,h^2\,+\,h^4\,+\,h^6\,=\,\ldots \Longrightarrow f_\infty\,=\,\frac{1}{1-h^2}
\ee
where $h^2=\frac 14 \,h_{++}h_{--}$.  Making the same procedure for $A_-$ and using (\ref{37}) and (\ref{42}),
\begin{equation}
\l{51}
A^i_{\pm}\,=\,{\frac{1 }{{1-h^2}}}\,\left[\,{\frac{1 }{2}} J_{\pm}^i\,-\,{\frac{1 }{%
4}}h_{++}J_{\mp}^i\,\right] \;\;.
\end{equation}
Hence, bringing these results back in (\ref{lag})  we have finally that,
\be
\l{52}
{\cal L} \,=\, \frac{1+h^2}{1-h^2}\, \partial_+ \Phi^i \partial_- \Phi^i\,+\,{%
\frac{h_{--} }{2{(1-h^2)}}} \partial_+ \Phi^i \partial_+ \Phi^i \,+\,{\frac{h_{++} }{2{(1-h^2)}}} \partial_- \Phi^i \partial_- \Phi^i\;\;,
\ee
where $\Phi^i=\phi^i-\rho^i$.  In other words
\begin{equation}
\label{alleqmovs}
{\cal L}\,=\,{\frac{1 }{2}}\,\sqrt{-g}\,g^{\mu\nu}\, \partial_{\mu} \Phi\,
\partial_{\nu} \Phi
\end{equation}
where
\be
\sqrt{-g}\,g^{\mu\nu}\,=\;\frac{1}{1-h^2}
\left( \begin{array}{cc}h_{--} & 1+h^2 \\
1+h^2 & h_{++}
\end{array}  \right)
\ee
and the metric has been modified by a constructive interference phenomenon.

To promote a profound investigation in this constructive interference, let us consider a
perturbative solution for this problem. To this end let us write (\ref{eqmova}) and (\ref{eqmovb}) as,
\begin{eqnarray}  
\label{A10}
A^i_+\,&=&\,{\frac{1 }{2}} J_+^i\,-\,{\frac{1 }{4}}h_{++}\,J_-^i \,+\,{\frac{%
1 }{4}}h^2\,A^i_+  \nonumber \\
A^i_-\,&=&\,{\frac{1 }{2}} J_-^i\,-\,{\frac{1 }{4}}h_{--}\,J_+^i \,-\,{\frac{%
1 }{4}}h^2\,A^i_-
\end{eqnarray}
and consider the weak field approximation (WFA) where terms of $O(h^2)\rightarrow 0
$. Notice that (\ref{A10}) gives the same result as (\ref{51}).
To simplify the notation we introduce the vector in the internal space as
${\ \phi}=\phi^i\,{\hat e}_i\,$, ${\bf A}_{\pm}=A^i_{\pm}{\hat e}_i\,$, etc..
Expanding these equations in powers of $h^2$ we have, in the zeroth order
approximation,
\begin{eqnarray}
\l{56}
{\bf A^{(0)}_+}\,&=&\,\partial_+ \rho\,+\,{\frac{1 }{2}} h_{++}
(\,\partial_- \rho\,-\,\partial_- \phi\,)\;\;  \nonumber \\
{\bf A^{(0)}_-}\,&=&\,\partial_+ \phi\,-\,{\frac{1 }{2}} h_{--}
(\,\partial_+ \rho\,-\,\partial_+ \phi\,)\;\;,
\end{eqnarray}
The Lagrangian (\ref{lag}) and the Noether currents are respectively
\be  
\label{lag2}
{\cal L}\,=\,{\cal L}_+^0\,+\,{\cal L}_-^0\,-\,{\bf A}_+\,{\bf J}_-\,-\,{\bf %
A}_-\,{\bf J}_+\,+\,{\frac{h_{--} }{2}}\,({\bf A}_+)^2\, +\,{\frac{h_{++} }{2}%
}\,({\bf A}_-)^2\,+\,2\,{\bf A}_-\,{\bf A}_+
\ee
and
\begin{eqnarray}
J^+_i\,&=&\,2\,\partial_- \phi^i\,+\,h_{--} \partial_+ \phi^i \\
J^-_i\,&=&\,2\,\partial_+ \rho^i\,+\,h_{++} \partial_- \rho^i
\end{eqnarray}

Substituting all the values in (\ref{lag2})
\be
\l{60}
{\cal L}_{WFA}\,=\,\partial_+ \Phi^i \partial_- \Phi^i\,+\,{\frac{h_{--} }{2}}
\partial_+ \Phi^i \partial_+ \Phi^i \,+\,{\frac{h_{++} }{2}} \partial_-
\Phi^i \partial_- \Phi^i\;\;,
\ee
where, as usual, $\Phi^i=\phi^i-\rho^i$.
Considering the $h^2$ term in the $W_2$ ($d_{ijk} \rightarrow 0$) in
\begin{equation}
A^i_{\pm}\,=\,{\frac{1 }{{1-h^2}}}\,\left[\,{\frac{1 }{2}} J^{\mp}_i\,-\,{\frac{1}{%
4}}h_{\pm\pm}J^{\pm}_i\,\right] \;\;,
\end{equation}
where $h^2=\frac 14 h_{++}h_{--}$.  

So, in (\ref{lag2}) we have,
\be
\l{62}
{\cal L}_{h^2} \,=\, \frac{1+h^2}{1-h^2}\, \partial_+ \Phi^i \partial_- \Phi^i\,+\,{%
\frac{h_{--} }{2{(1-h^2)}}} \partial_+ \Phi^i \partial_+ \Phi^i \,+\,{\frac{h_{++}}{2{(1-h^2)}}} \,\partial_- \Phi^i \partial_- \Phi^i
\ee
or
\begin{equation}
{\cal L}_{h^2}\,=\,{\frac{1 }{2}}\,\sqrt{-g}\,g^{\mu\nu}_{(0)}\, \partial_{\mu}
\Phi\, \partial_{\nu} \Phi
\end{equation}
which is the result obtained from (\ref{alleqmovs}) when
$\sqrt{-g}\,g^{\mu\nu}\stackrel{h^2\to 0}{\longrightarrow} \sqrt{-g^{}_{(0)}}\,g^{\mu\nu}_{(0)}$.

We can see clearly that (\ref{60}) is the zeroth order approximation of the action (\ref{52}) with $h^2 \rightarrow 0$, i.e.,
\be
{\cal L}_{WFA}\,=\,{\cal L}_{h^2}\,(h^2 \rightarrow 0)\;\;.
\ee
Hence we can assume that the perturbative procedure in the soldering fields have disclosed an interesting behavior.  The zeroth order approximation written in (\ref{56}) showed that in (\ref{60}) the interference between the two distinct chiralities in ${\cal W}_2$ is, unless a cross term, the simple sum of both actions.  However, considering the $h^2$ terms we see that it is not true.
As we have stressed, (59) is not a trivial result: both chiral particles are now parts of the same multiplet and we have the modification of the metric through an constructive interference.
We can see clearly that (57) is just a $h^2 \rightarrow 0$ approximation of the exact action written in (59) or compactly written in (60).
Next we will prove in a precise way that this behavior can be seen in all spin-s ${\cal W}$-theories.

\subsection{Weak field approach to the soldering of chiral ${\cal W}_3$}

Let us next analyze the ${\cal W}_3$ model for the right-handed chirality,
equation (\ref{oito}),
\be
{\cal L}_+^0 \,=\, \partial_+ \phi^i \partial_- \phi^i\,+\,{\frac{h_{--} }{2}}
\partial_+ \phi^i \partial_+ \phi^i \,+\,\,{\frac{B_{---} }{3}} d_{ijk}
\partial_+ \phi^i \partial_+ \phi^j \partial_+ \phi^k\;\;,
\ee
this is the action for lowest nonminimal coupling \cite{ssn2} with $k$, the expansion parameter for the Noether method equal to $-1$.

The gauge transformation is
\begin{equation}
\phi^i \rightarrow \phi^i \,+\,\alpha^i
\end{equation}

The gauge variation of the model is
\begin{equation}
\delta {\cal L}_+^0\,=\,J^+_i \partial_+ \alpha^i\;\;,
\end{equation}
where $J^+_i$ is the left Noether current
\begin{equation}
J^+_i\,=\,2\,\partial_- \phi^i\,+\,h_{--} \partial_+ \phi^i\,+\,B_{---}
d_{ijk} \partial_+ \phi^j \partial_+ \phi^k
\end{equation}

Following again the soldering algorithm, we have after two iterations that
\begin{equation}
\delta {\cal L}^2_+\,=\,-\,2\,A_+^i \delta A_-^i\;\;,
\end{equation}
where $A^i_{\pm}$ are the soldering fields.

For the left chirality we can write
\be
{\cal L}^0_- \,=\, \partial_+ \rho^i \partial_- \rho^i\,+\,{\frac{h_{++} }{2}}
\partial_- \rho^i \partial_- \rho^i \,+\,{\frac{B_{+++} }{3}} d_{ijk}
\partial_- \rho^i \partial_- \rho^j \partial_- \rho^k
\ee

The gauge variation of the model is
\begin{equation}
\delta {\cal L}_-^0\,=\,J^-_i \partial_- \alpha^i\;\;,
\end{equation}
where $J^-_i$ is the right Noether current
\begin{equation}
J^-_i\,=\,2\,\partial_+ \rho^i\,+\,h_{++} \partial_- \rho^i\,+\,B_{+++}
d_{ijk} \partial_- \phi^j \partial_- \phi^k
\end{equation}

After two iterations we have that
\begin{equation}
\delta {\cal L}^2_-\,=\,-\,2\,A_-^i \delta A_+^i\;\;,
\end{equation}
and the final soldered action is
\begin{equation}
{\cal L}_{FINAL}\,=\,{\cal L}^2_+\,+\,{\cal L}^2_-\,+\,2\,A^i_-\,A^i_+
\end{equation}
which has a vectorial gauge invariance, i.e., $\delta {\cal L}\,=\,0$.

Substituting all the ${\cal L}_\pm^{(N)}$ we can write explicitly the final form of the action as being
\begin{eqnarray}  
\label{lagg}
{\cal L}_{FINAL}\,&=&\,{\cal L}_+^0\,+\,{\cal L}_-^0\,-\,A^i_+\,J^+_i\,-\,A^i_-\,J^-_ i\, 
+\,{\frac{h_{--} }{2}}\,(A^i_+)^2\,+\,{\frac{h_{++} }{2}}\,(A^i_-)^2  \nonumber \\
&+&\,B_{---} d_{ijk}[\, A^i_+\,A^j_+\, \partial_+ \phi^k\,-\,{\frac{1 }{3}}%
A^i_+ A^j_+ A^k_+\,] \nonumber \\
&+&B_{+++} d_{ijk}[\, A^i_-\,A^j_- \,\partial_- \rho^k\,-\,{\frac{1 }{3}}%
A^i_- A^j_-A^k_-\,]\, \nonumber \\
&+&\,2\,A^i_-\,A^i_+
\end{eqnarray}

Solving the equations of motion for the soldering fields, these can be eliminated
in favor of the other variables
\be
\l{eqmova2}
A^i_+\,=\,{\frac{1 }{2}} J^-_i\,-\,{\frac{1 }{2}}h_{++}\,A^i_-\,+\,B_{+++} d_{ijk}\, A^j_-\,(\,{\frac{A^k_- }{6}}\,-\,{\frac{\partial_+ \rho^k
}{2}}), 
\ee
\be 
\l{eqmovb2}
A^i_-\,=\,{\frac{1 }{2}} J^+_i\,-\,{\frac{1 }{2}}h_{--}\,A^i_+\,+\,B_{---} d_{ijk}\, A^j_+\,(\,{\frac{A^k_+ }{6}}\,-\,{\frac{\partial_+ \phi^k
}{2}}), 
\ee
Substituting (\ref{eqmovb2}) in (\ref{eqmova2}) and, to be concise, writing only 
the solution for the $A^i_+$ we have
\begin{eqnarray}  \label{sol}
 A^i_+\,&=&\,{\frac{1 }{2}} J^-_i\,-\,{\frac{1 }{4}}h_{++}\,J_i^+\,+\,{\frac{%
1 }{24}} B_{+++} d_{ijk}\, J_j^+\,J_k^+\,-\,{\frac{1 }{4}} B_{+++} d_{ijk}\,
J_j^+\,J_k^+ \partial_- \rho^k \,+\,{\frac{1 }{4}} h_{++}h_{--}\,A^i_+\,\nonumber \\
&+&\,{\frac{1 }{4}}h_{++}B_{---}
d_{ijk}\, A^j_+\,\partial_+ \phi^k \,-\,{\frac{1 }{12}}h_{--}B_{+++}
d_{ijk}\, J^+_j A^k_+ \,-\,{\frac{1}{12}}B^2 d_{ijk}d_{kmn}\, J^+_j A^m_+\, \partial_+ \phi^n \nonumber \\
&+&\,{\frac{1}{4}}h_{--}B_{+++}d_{ijk} A^j_+ \partial_- \rho^k\, +\,{\frac{1}{4}}B^2d_{ijk}d_{jmn}A^m_+ \partial_+ \phi^n \partial_- \rho^k\,-\,{\frac{1}{12}} h_{++}B_{---} d_{ijk} A^j_+ A^k_+  \nonumber \\
&+&\,{\frac{1}{36}} d_{ijk}d_{kmn} J^+_j A^m_+ A^n_+\, 
+\,{\frac{1}{24}}
h^2_{--}B_{+++}d_{ijk}A^j_+ A^k_+ \,+\,{\frac{1}{12}}h_{--}B^2d_{ijk}d_{kmn}A^j_+ A^m_+\partial_+ \phi^n\, \nonumber \\
&+&\,{\frac{1}{24}}B_{---}B^2d_{ijk}d_{jmn}d_{kpq}A^m_+ A^p_+\partial_+ \phi^n \partial_+ \phi^q\,-\,{\frac{1}{12}}B^2d_{ijk}d_{jmn}A^m_+ A^n_+\partial_- \rho^k  \nonumber \\
&-&\,{\frac{1}{36}}h_{--}B^2 d_{ijk}d_{jmn}\, A^k_+ A^m_+ A^n_+\, \nonumber \\
&-&\,{\frac{1}{72}}\,B_{---}B^2 d_{ijk}d_{jmn}d_{kpq}\, A^m_+ A^p_+ A^n_+ 
(\,\partial_+ \phi^q A^n_+\,+\,\partial_+ \phi^n A^q_+\,)  \nonumber \\
&+&\,{\frac{1 }{216}}B_{---}B^2 d_{ijk}d_{jmn}d_{kpq}\,
A^m_+A^p_+A^n_+A_+^q\;\;,
\end{eqnarray}
where $B^2=B_{+++}B_{---}$.

Now we can solder the Lagrangian with $d_{ijk} \neq 0$ and take the terms of
zero order ($h^2 \rightarrow 0$) of $A^i_+$ in (\ref{sol}), with the Noether currents,
\bn
A^i_+\,&=&\,\partial_+ \rho^i\,+\,{\frac{1 }{2}}h_{++} (\,\partial_-
\rho^i\,-\,\partial_- \phi^i\,)\,+\,{\frac{1}{2}}B_{+++}d_{ijk}\,\partial_-
\rho^k\,(\,\partial_- \rho^j\,-\,\partial_- \phi^j\,)\, \nonumber \\
&+&\,{\frac{1 }{6}}B_{+++}d_{ijk} \partial_- \phi^j \partial_- \phi^k, \l{eqa}
\en
\bn
A^i_-\,&=&\,\partial_- \phi^i\,+\,{\frac{1 }{2}}h_{--} (\,\partial_+
\rho^i\,-\,\partial_+ \phi^i\,)\,+\,{\frac{1 }{2}}B_{---}d_{ijk}\,\partial_+
\phi^k\,(\,\partial_+ \phi^j\,-\,\partial_- \rho^j\,)\, \nonumber \\
&+&\,{\frac{1 }{6}}B_{---}d_{ijk} \partial_+ \rho^j \partial_+ \rho^k, \l{eqb}
\en

Substituting in (\ref{lagg}) and computing the soldering fields through the
equations of motion we have the soldered action for the WFA action
and after an arduous algebra we have considered the $h^2$ terms, with $\Phi^i=\phi^i-\rho^i$,
\begin{eqnarray}
{\cal L}_{WFA}\,&=&\,\partial_+ \Phi^i \partial_- \Phi^i\,+\,\frac 12 \,h_{--} \partial_+
\Phi^i \partial_+ \Phi^i \,+\,\frac 12 \,h_{++} \partial_- \Phi^i \partial_- \Phi^i\,+\,{\frac{1 }{3}}B_{+++} d_{ijk} \partial_- \Phi^i \partial_- \Phi^j \partial_- \Phi^k \, \nonumber \\
&+&\,{\frac{1 }{3}}B_{---} d_{ijk} \partial_+ \Phi^i
\partial_+ \Phi^j \partial_+ \Phi^k 
\l{lagfinala}
\end{eqnarray}
\bn
{\cal L}_{h^2} &=& \frac{1+h^2}{1-h^2}\, \partial_+ \Phi^i \partial_- \Phi^i\,+\,{\frac{h_{--} }{2{(1-h^2)}}} \partial_+ \Phi^i \partial_+ \Phi^i \,+\,{\frac{h_{++} }{2{(1-h^2)}}} \partial_- \Phi^i \partial_- \Phi^i \nonumber \\
&+&\,{\frac{1 }{3(1-h^2)}}B_{+++} d_{ijk} \partial_- \Phi^i \partial_- \Phi^j
\partial_- \Phi^k \,
+\,{\frac{1 }{3(1-h^2)}}B_{---} d_{ijk} \partial_+ \Phi^i
\partial_+ \Phi^j \partial_+ \Phi^k \nonumber \\
& &
\l{lagfinalb}
\end{eqnarray}
where we can see again that
\be
{\cal L}_{WFA}\,=\,{\cal L}_{h^2}\,(h^2 \rightarrow 0)\;\;,
\ee
demonstrating that what occurred to the ${\cal W}_2$-theory happened to ${\cal W}_3$ and again we have a constructive interference modifying the metric.

This first-order action (\ref{lagfinala}) is similar to that found out by Schoutens {\it et al} \cite{ssn2} for the spin-s theory (the SSN action), to describe a ${\cal W}$-string propagating on a flat background spacetime metric. The $\partial_{\pm}\Phi^i$ substitutes the so-called ``nested covariant derivatives".
Besides we have also obtained a reduction in the infinite nonlinearity.
The soldered action couples both chiral
scalar fields to a dynamical gauge field.  This action is characteristic of the interference
process which leads to a new and nontrivial result as the modification of the metric by a constructive interference and proves that the SSN action can be an approximation of a more general action.  Next we will see the spin-s$> 3$ generalization of the SSN-like action.

\subsection{${\cal W}_N$, $N \geq 4$}

In this section we will write only the final results of the interference mechanism, i.e., 
for the ${\cal W}_4$ algebra as we know,
\bn
{\cal L}_+^0 \,&=&\, \partial_+ \phi^i \partial_- \phi^i\,+\,{\frac{h_{--}}{2}}
\partial_+ \phi^i \partial_+ \phi^i \,+\,{\frac{B_{---}}{3}} d_{ijk}
\partial_+ \phi^i \partial_+ \phi^j \partial_+ \phi^k \,  \nonumber \\
&+&\,{\frac{B_{----}}{4}} d_{ijkl}
\partial_+ \phi^i \partial_+ \phi^j \partial_+ \phi^k \partial_+ \phi^l \l{W4a}
\en
\bn
{\cal L}^0_- \,&=&\, \partial_+ \rho^i \partial_- \rho^i\,+\,{\frac{h_{++}}{2}}
\partial_- \rho^i \partial_- \rho^i \,+\,{\frac{B_{+++} }{3}} d_{ijk}
\partial_- \rho^i \partial_- \rho^j \partial_- \rho^k\, \nonumber \\
&+&\,{\frac{B_{++++}}{4}} d_{ijkl}
\partial_- \rho^i \partial_- \rho^j \partial_- \rho^k \partial_- \rho^l\;\;, 
\l{W4b}
\en
where $d_{ijkl}$ is another symmetric tensor \cite{hull2}.

After all the interference method, the WFA and the $h^2$-term actions are respectively
\begin{eqnarray}
{\cal L}_{WFA}&=&\,\partial_+ \Phi^i \partial_- \Phi^i+\frac 12 \,h_{--} \partial_+
\Phi^i \partial_+ \Phi^i \,+\frac 12 \,h_{++} \partial_- \Phi^i \partial_- \Phi^i+ 
\,+\,{\frac{1 }{3}}B_{+++} d_{ijk} \partial_- \Phi^i \partial_- \Phi^j
\partial_- \Phi^k \, \nonumber \\
&+&{\frac{1 }{3}}B_{---} d_{ijk} \partial_+ \Phi^i
\partial_+ \Phi^j \partial_+ \Phi^k\,+\,{\frac{1 }{4}}B_{++++} d_{ijkl} \partial_- \Phi^i \partial_- \Phi^j
\partial_- \Phi^k \partial_- \Phi^l\, \nonumber \\
&+&\,{\frac{1 }{4}}B_{----} d_{ijkl} \partial_+ \Phi^i
\partial_+ \Phi^j \partial_+ \Phi^k \partial_+ \Phi^l  \l{lagfina}
\en
\bn
{\cal L}_{h^2} &=& \frac{1+h^2}{1-h^2}\, \partial_+ \Phi^i \partial_- \Phi^i + {
\frac{h_{--} }{2{(1-h^2)}}} \partial_+ \Phi^i \partial_+ \Phi^i \,
+\,{\frac{%
h_{++} }{2{(1-h^2)}}} \partial_- \Phi^i \partial_- \Phi^i \nonumber \\
&+&{\frac{1 }{3(1-h^2)}}B_{+++} d_{ijk} \partial_- \Phi^i \partial_- \Phi^j
\partial_- \Phi^k \,+\,{\frac{1 }{3(1-h^2)}}B_{---} d_{ijk} \partial_+ \Phi^i\partial_+ \Phi^j \partial_+ \Phi^k \nonumber \\
&+&\frac{B_{++++}}{4(1-h^2)} d_{ijkl} \partial_- \Phi^i \partial_- \Phi^j
\partial_- \Phi^k \partial_- \Phi^l 
\,+\,\frac{B_{----}}{4(1-h^2)}
d_{ijkl} \partial_+ \Phi^i \partial_+ \Phi^j \partial_+ \Phi^k \partial_+ \Phi^l \nonumber \\
& &
\l{lagfinb}
\end{eqnarray}

Finally, for a ${\cal W}$-gravity of spin-s, for both chiralities respectively,
\bn
{\cal L}_+^0 &=& \partial_+ \phi^i \partial_- \phi^i+{\frac{h_{--} }{2}}
\partial_+ \phi^i \partial_+ \phi^i \,+\,{\frac{B_{---} }{3}} d_{ijk}
\partial_+ \phi^i \partial_+ \phi^j \partial_+ \phi^k \nonumber \\
&+&\frac 14 \,B_{----} d_{ijkl}
\partial_+ \phi^i \partial_+ \phi^j \partial_+ \phi^k \partial_+ \phi^l\,
+\, \ldots +\frac{1}{s} B_{--\ldots-} d_{i_1 i_2 \ldots i_s} \partial_+ \phi^{i_1}
\partial_+ \phi^{i_2} \ldots \partial_+ \phi^{i_s} \nonumber \\
& &
\l{spina}
\en
\bn
{\cal L}^0_- &=& \partial_+ \rho^i \partial_- \rho^i+{\frac{h_{++}}{2}}
\partial_- \rho^i \partial_- \rho^i \,+\,{\frac{B_{+++} }{3}} d_{ijk}
\partial_- \rho^i \partial_- \rho^j \partial_- \rho^k \nonumber \\
&+&\frac 14 \,B_{++++} d_{ijkl}
\partial_- \rho^i \partial_- \rho^j \partial_- \rho^k \partial_- \rho^l \,+\, \ldots + \frac 1s B_{++\ldots+} d_{i_1 i_2 \ldots i_s} \partial_- \phi^{i_1}
\partial_- \phi^{i_2} \ldots \partial_- \phi^{i_s} \nonumber \\
& & 
\l{spinb}
\en

\ni and the final actions are
\begin{eqnarray}
{\cal L}_{WFA} &=& \,\partial_+ \Phi^i \partial_- \Phi^i+\frac 12 \,h_{--} \partial_+
\Phi^i \partial_+ \Phi^i \,+\,\frac 12 \,h_{++} \partial_- \Phi^i \partial_- \Phi^i+ \nonumber \\
&+&{\frac{1 }{3}}B_{+++} d_{ijk} \partial_- \Phi^i \partial_- \Phi^j
\partial_- \Phi^k \,+\,{\frac{1 }{3}}B_{---} d_{ijk} \partial_+ \Phi^i
\partial_+ \Phi^j \partial_+ \Phi^k  \nonumber \\
&+&{\frac{1 }{4}}B_{++++} d_{ijkl} \partial_- \Phi^i \partial_- \Phi^j
\partial_- \Phi^k \partial_- \Phi^l \,+\,{\frac{1 }{4}}B_{----} d_{ijkl} \partial_+ \Phi^i
\partial_+ \Phi^j \partial_+ \Phi^k \partial_+ \Phi^l \nonumber \\
&+& \ldots \,
+\,\frac{B_{++\ldots+}}{s} d_{i_1 i_2 \ldots i_s} \partial_- \Phi^{i_1}
\partial_- \Phi^{i_2} \ldots \partial_- \Phi^{i_s}\,
+\,\frac{B_{--\ldots-}}{s} d_{i_1 i_2 \ldots i_s} \partial_+ \Phi^{i_1}
\partial_+ \Phi^{i_2} \ldots \partial_+ \Phi^{i_s} \nonumber \\
& &
\label{lagfinalspina}
\en
\bn
{\cal L}_{h^2} &=& \frac{1+h^2}{1-h^2}\, \partial_+ \Phi^i \partial_- \Phi^i+{\frac{h_{--} }{2{(1-h^2)}}} \partial_+ \Phi^i \partial_+ \Phi^i \, 
+\,{\frac{h_{++} }{2{(1-h^2)}}} \partial_- \Phi^i \partial_- \Phi^i \nonumber \\
&+&{\frac{1}{3(1-h^2)}}B_{+++} d_{ijk} \partial_- \Phi^i \partial_- \Phi^j
\partial_- \Phi^k \,+\,{\frac{1 }{3(1-h^2)}}B_{---} d_{ijk} \partial_+ \Phi^i
\partial_+ \Phi^j \partial_+ \Phi^k \nonumber \\
&+&{\frac{1 }{4(1-h^2)}}B_{++++} d_{ijkl} \partial_- \Phi^i \partial_- \Phi^j
\partial_- \Phi^k \partial_- \Phi^l\,
+\,{\frac{1}{4(1-h^2)}}B_{----}
d_{ijkl} \partial_+ \Phi^i \partial_+ \Phi^j \partial_+ \Phi^k \partial_+ \Phi^l \nonumber \\
&+& \ldots \,+\, \frac{B_{++\ldots+}}{s(1-h^2)} d_{i_1 i_2 \ldots i_s} \partial_- \Phi^{i_1}
\partial_- \Phi^{i_2} \ldots \partial_- \Phi^{i_s} 
\,+\, \frac{B_{--\ldots-}}{s(1-h^2)} d_{i_1 i_2 \ldots i_s} \partial_+ \Phi^{i_1}
\partial_+ \Phi^{i_2} \ldots \partial_+ \Phi^{i_s} \nonumber \\
& &
\label{lagfinalspinb}
\end{eqnarray}

Now we have a first-order action for spin-$s\geq 3$ theories.  Hence, we can conjecture if the SSN action is a $h^2\rightarrow 0$ approximation of a more general action as can be seen comparing (\ref{lagfinalspina}) and (\ref{lagfinalspinb}).

\section{Final remarks and perspectives}

The quantization of such a system of matter coupled to gravity defines a string theory.  This interesting behavior warrant a study of fusion of W-algebra coupled to gravity.  We have obtained an action similar to that obtained by Schoutens {\it et al} for spin-3 gravities.  The result showed us that the SSN action can be an approximation of a more general action where the metric is modified.  We have demonstrated in a precise way that this behavior is confirmed in spin-s$> 3$ gravities.

As a final remark, in particular for further studies, we can analyze the chiral $W_3$ gravity, where, to cancel the anomaly we have to add finite local counterterms.  Considering the non-chiral $W_3$ gravity, it can be analyzed the relation between the dynamical (chiral) decomposition and the factorization that occurs in a closed $W_3$ string.  There, the Hilbert space factorizes as usual into a tensor product of the Hilbert space of the left-moving states with the one of the right-moving states.  The Hilbert space $H$ of the left-movers is then the product of the single-boson Fock space $F_{\phi}$ with the Hilbert space of the effective conformal field theory, $\tilde{H}$, and $H=F_{\phi} \otimes \tilde{H}$.

\appendix
\section{A review of the interference mechanism}

The technique of soldering (interference) essentially comprises in simultaneously lifting the gauging of a
global symmetry of a couple of self-dual and antiself-dual actions
to their local version \cite{ms,cw}.
We remark that the direct sum of the
classical actions depending on different fields would not give anything new.
It is the soldering process that leads to a new and nontrivial result.

The basic idea of the soldering procedure is to raise a global Noether
symmetry of the constituents actions into a local one, but
for an effective composite system, consisting of the dual components and an
interference term. This algorithm, consequently, defines the soldered
action. Here we shall adopt an iterative Noether procedure to lift the
global symmetries. Therefore, assume that the symmetries in question are
being described by the local actions $S_{\pm }(\phi _{\pm }^{\eta })$,
invariant under a global multi-parametric transformation
\begin{equation}
\delta \phi _{\pm }^{\eta }=\alpha ^{\eta }  \label{ii10}
\end{equation}

Here $\eta$ represents the tensorial character of the basic fields in the
dual actions $S_{\pm }$ and, for notational simplicity, will be dropped from
now on. Now, under local transformations these actions will not remain
invariant, and Noether counter-terms become necessary to reestablish the
invariance, along with appropriate compensatory soldering fields $B^{(N)}$,
\be
S_{\pm }(\phi _{\pm })^{(0)} \rightarrow S_{\pm }(\phi _{\pm })^{(N)}=S_{\pm
}(\phi _{\pm })^{(N-1)}-B^{(N)}J_{\pm }^{(N)}  \label{ii20}
\ee

Here $J_{\pm }^{(N)}$ are the Noether currents, and $N$ is the iteration
number. For the self and antiself-dual systems we have in mind that this
iterative gauging procedure is (intentionally) constructed not to produce
invariant actions for any finite number of steps. However, if after $N$
repetitions the non invariant piece ends up being only dependent on the
gauging parameters, but not on the original fields, there will exist the
possibility of mutual cancelation, if both self and anti-self gauged systems
are put together. Then, suppose that after N repetitions we arrive at the
following simultaneous conditions,
\begin{eqnarray}  \label{ii30}
\delta S_{\pm}(\phi_{\pm})^{(N)} &\neq& 0  \nonumber \\
\delta S_{B}(\phi_{\pm})&=& 0
\end{eqnarray}
with
\begin{equation}  \label{ii40}
S_{B}(\phi_{\pm})=S_{+}^{(N)}(\phi_{+}) + S_{-}^{(N)}(\phi_{-})+ %
\mbox{Contact Terms}\;\;,
\end{equation}
then we can immediately identify the (soldering) interference term as,
\begin{equation}  \label{ii50}
S_{int}=\mbox{Contact Terms}-\sum_{N}B^{(N)} J_{\pm}^{(N)}\;\;,
\end{equation}
where the Contact Terms are generally higher order functions of the
soldering fields. Incidentally, these auxiliary fields $B^{(N)}$ may be
eliminated, in some cases, from the resulting effective action in favor of
the physically relevant degrees of freedom. It is important to notice that
after elimination of the soldering fields, the resulting effective action
will not depend on either self or anti-self dual fields $\phi_{\pm}$ but
only in some collective field, say $\Phi$, defined in terms of the original
ones in a (Noether) invariant way
\begin{equation}
S_{B}(\phi _{\pm })\rightarrow S_{eff}(\Phi ).  \label{ii60}
\end{equation}
Once such effective action has been established, the physical consequences
of the soldering are readily obtained by simple inspection. This will
progressively be clarified in the specific applications, the ${\cal W}$-theories, to be given in the sections that follow.

\section{Acknowledgments}

E.M.C.A. is financially supported by Funda\c{c}\~ao de Amparo \`a Pesquisa
do Estado de S\~ao Paulo (FAPESP). This work was also
partially supported by Conselho Nacional de Pesquisas e Desenvolvimento
(CNPq).  C. W. would like to thank the hospitality of the Departamento de F\'{\i}sica e Qu\'{\i}mica from UNESP/Guaratinguet\'a, where part of this work was made and FAPESP for financial support. FAPESP and CNPq are brazilian research agencies.

\end{document}